\documentclass[%
 reprint,
 amsmath,amssymb,
 aps,
prb,
showpacs
]{revtex4-2}
\pdfoutput=1
\usepackage{graphicx}
\usepackage{dcolumn}
\usepackage{bm}
\usepackage{xcolor}
\usepackage{amsmath,amsfonts,amssymb}
\usepackage{mathtools}
\usepackage[font=footnotesize,labelfont=bf]{caption}
\usepackage[font=footnotesize,labelfont=bf]{subcaption}
\begin{document}

\preprint{APS/123-QED}

\title{Elastic anisotropy in heterogeneous materials} 

\author{Abhilash M Nagaraja}
\email{abhilashn@iisc.ac.in ;}
\affiliation{Department of Aerospace Engineering, Indian Institute of Science, Bangalore 560012, India.}


\begin{abstract}
Heterogeneous materials exhibit anisotropy which is influenced by factors such as individual phase properties and microstructural configuration that form crucial descriptors of heterogeneity. A review of anisotropy indices proposed in the context of single crystals reveals that they do not account for the descriptors of heterogeneity limiting their utility in heterogeneous materials. To address this shortcoming, this work presents a novel approach to quantify elastic anisotropy in heterogeneous materials utilizing their effective elastic properties. Anisotropy quantification has been demonstrated considering two phase composite materials highlighting the crucial role of constituent volume fractions, secondary phase shape, and elastic contrast in influencing the extent of anisotropy. Additionally, specific material and microstructural descriptors leading to overall isotropy in composite materials are validated and an extension of the approach to account for defects such as porosity and cracks is presented. The presented results demonstrate the applicability of the anisotropy quantification approach to any configuration of heterogeneous materials.
\end{abstract}

\maketitle


\section{Introduction}
Anisotropy, defined as the direction dependence of material response, is a key aspect characterizing material behavior. Various classes of natural and engineered materials such as single crystals \cite{ingel1988elastic}, polycrystals \cite{brenner2009elastic,zubelewicz2021another}, composite materials \cite{hashin1983}, biological tissues \cite{katz1987elastic,murphy2014evolution}, geological materials \cite{sondergeld2011}, 2D materials \cite{li2020elastic} etc. exhibit anisotropy to varying degree. Additionally, anisotropy plays an important role in crucial material phenomena such as phase transformations \cite{SCHMITZ1989}, material damage \cite{fichant1997}, dislocation dynamics \cite{Han2003}, defect mobility \cite{jan2008}, fracture behavior \cite{tvergaard1988}, plastic deformation in crystals \cite{li2002} etc. The pervasive nature of anisotropy illustrates the importance of anisotropy quantification by relevant measures.

Although elastic anisotropy type can be represented graphically \cite{nordmann2018visualising,ran2023velas} and by material symmetry groups \cite{coleman1964material}, a quantifiable parameter in the form of a scalar anisotropy index facilitates the comparison of anisotropy within and across material symmetry groups. The extent of elastic anisotropy can be evaluated from the elastic behavior of materials represented by generalized Hooke's law ($\sigma_{ij}= C_{ijkl} \epsilon_{kl} $) consisting of a fourth-rank elastic stiffness tensor ($C_{ijkl}$) relating the second rank stress ($\sigma_{ij}$) and strain ($\epsilon_{kl}$) tensors that can be represented in matrix notation through Voigt contraction in the following form.

\begin{equation}
\label{eq:cijkl}
  {
   \begin{bmatrix}
   \sigma_{11} \\ \sigma_{22}  \\ \sigma_{33} \\ \sigma_{23} \\ \sigma_{13} \\ \sigma_{12} \\
 \end{bmatrix}  = 
 \begin{bmatrix}
   C_{11} &  C_{12} & C_{13} &  C_{14} & C_{15} & C_{16} \\
   C_{21} &  C_{22} & C_{23} &  C_{24} & C_{25} & C_{26} \\
   C_{31} &  C_{32} & C_{33} &  C_{34} & C_{35} & C_{36} \\
   C_{41} &  C_{42} & C_{43} &  C_{44} & C_{45} & C_{46} \\
   C_{51} &  C_{52} & C_{53} &  C_{54} & C_{55} & C_{56} \\
   C_{61} &  C_{62} & C_{63} &  C_{64} & C_{65} & C_{66} \\
 \end{bmatrix} 
 \begin{bmatrix}
   \epsilon_{11} \\ \epsilon_{22}  \\ \epsilon_{33} \\ 2\epsilon_{23} \\ 2\epsilon_{13} \\ 2\epsilon_{12} \\
 \end{bmatrix} 
 } 
\end{equation}

Anisotropy quantification in single crystals has been explored by a diverse range of anisotropy indices. Zener's anisotropy index $A^{Z}=2 C_{44}/(C_{11}-C_{12})$, where $C_{11}$, $C_{12}$, and $C_{44}$ are the independent single crystal elastic constants \cite{zener1948}, quantifies anisotropy as the ratio of the extreme values of orientation dependent shear moduli that is valid for crystals with cubic symmetry. However, $A^{Z}$ measures anisotropy as deviation from unity that indicates isotropy making it ambiguous since values both lower and higher than unity indicate anisotropy.

Chung and Buessem \cite{chung1967elastic} proposed a measure of anisotropy based on the equality of shear moduli obtained from Voigt ($\mu^V$) and Reuss average ($\mu^R$) of shear modulus leading to an anisotropy index in the form $A^{C}=(\mu^V-\mu^R)/(\mu^V+\mu^R)$. Although $A^{C}$ can be calculated for any crystal symmetry, disregarding the bulk modulus that influences anisotropy of crystals other those exhibiting cubic symmetry, reduces it's utility in crystals with lower symmetries. Ledbetter and Migliori \cite{ledbetter2006} proposed a more general anisotropy index based on shear wave velocities in the form $A^{*}=\nu_2^2/\nu_1^2$ where $\nu_1$ and $\nu_2$ are respectively the minimum and maximum shear wave velocities among all propagation and polarization directions.

The above-mentioned anisotropy indices lack universality that stems from disregarding the tensorial nature of elastic stiffness as highlighted by Ranganathan and Ostoja-Starzewski \cite{ranganathan2008}. The implication of disregarding the tensorial nature of the elastic stiffness is that the bulk part of the stiffness tensor is neglected limiting the utility of the above anisotropy indices to crystals with isotropic bulk resistance. Anisotropy quantification considering the tensorial nature of elastic stiffness has been attempted via tensor norm based measures by Nye \cite{nye1985physical}, Zheng \cite{wang2007extreme} ($A_{Nye-Zheng}=\left\|C - C^{iso}\right\|/\left\|C\right\|$, where $C - C^{iso}$ is the anisotropic part of C), Zhang \cite{rychlewski1989anisotropy}, Rychlewski \cite{rychlewski1984evaluation} ($A_{Zhang-Rychlewski} = \max \left\|R*C-
C\right\|/2\left\|C\right\|$, where $R$ represents a rotation operation). In both the norm based measures, $\left\|C\right\|$ is the norm of a general fourth-rank tensor $C$ denoted as $\left\|C\right\|=(C_{ijkl}C_{ijkl})^{1/2}$.

A set of criteria for generality, universality, and applicability of an anisotropy index has been proposed by Fang et al. \cite{fang2019} that can be summarized in the following manner. An acceptable anisotropy index must (a) be zero in the case of isotropy with any positive number indicating anisotropy, (b) should approach infinity if there exists a direction along which the deformation resistance of the material approaches zero, and (c) must have a clear physical meaning. Of the anisotropy indices discussed previously, $A^Z$ and $A^*$ do not consider the tensorial nature of the elastic stiffness tensor and do not fulfill the first criteria for an acceptable anisotropy index. $A^C$ along with the tensor norm based measures $A_{Nye-Zheng}$ and $A_{Zhang-Rychlewski}$ do approach infinity if there exists a direction along which the deformation resistance of the material tends to zero thereby not satisfying the second criterion. These observations highlight the inadequacies of the discussed anisotropy indices proposed in the context of single crystals.

In the backdrop of the inadequacies of the above anisotropy indices, the following measures have been predominantly used in anisotropy quantification in single crystals.

(a) Universal anisotropy index $A^{U}$ that considers the tensorial nature of the stiffness tensor by relating anisotropy to the Euclidean distance between the Voigt ($C^V$) and Reuss bounds ($C^R$) of the stiffness tensor of a spatially uniformly distributed aggregate of a single crystal \cite{ranganathan2008}. $A^{U}$ is a scalar that can be expressed in terms of bulk ($\kappa$) and shear moduli ($\mu$) of a single crystal in the following form.

\begin{equation}
A^{U}(C^V,C^R)=\frac{\kappa^V}{\kappa^R}+5 \frac{\mu^V}{\mu^R}-6
\label{eq:anisotropy_ranganathan}
\end{equation}

Superscript $V$ and $R$ represent the Voigt and Reuss bounds obtained by ensemble average of randomly oriented crystals. Isotropic crystals have $A^U=0$ and any deviation from zero indicates anisotropy. 

(b) Log-Euclidean anisotropy index $A^L$ that relates anisotropy to the log-Euclidean distance between the Voigt ($C^V$) and Reuss bounds ($C^R$) of the stiffness tensor of a spatially uniformly distributed aggregate of a single crystal \cite{kube2016a} in the following form.

\begin{equation}
A^{L}(C^V,C^R)=\sqrt{\left[ln\left(\frac{\kappa^V}{\kappa^R}\right)\right]^2+5\left[ln \left(\frac{\mu^V}{\mu^R}\right)\right]^2}
\label{eq:anisotropy_kube}
\end{equation}

$A^L$ becomes zero for isotropic crystals and any deviation for zero indicates the extent of anisotropy that is valid for any crystal class. Additionally, $A^{L}$ being an absolute measure of anisotropy, the degree of anisotropy is comparable across crystals.

(c) Elastic strain energy based anisotropy index $A^{SE}$  defined as the ratio of maximum and minimum elastic strain energy across the possible range of strain combinations and crystal orientations \cite{fang2019}.

\begin{equation}
A^{SE}=\max_{\epsilon,R_{\epsilon}^{(1)},R_{\epsilon}^{(2)}}\left(\frac{(R_{\epsilon}^{(1)}\epsilon)^T C (R_{\epsilon}^{(1)}\epsilon)}{(R_{\epsilon}^{(2)}\epsilon)^T C (R_{\epsilon}^{(2)}\epsilon)}\right) -1 
\label{eq:ASE}
\end{equation}

Where $R_{\epsilon}$ is the rotation matrix of strain, $\epsilon$ is the strain tensor and $C$ is the stiffness tensor. $A^{SE}$ can be expressed in stress or strain space that are equivalent \cite{fang2019}.

Although $A^U$ and $A^L$ satisfy the first two criteria for an acceptable anisotropy index, they relate anisotropy to the distance between Voigt and Reuss bounds (that converge when a material is isotropic) making their physical meaning difficult to comprehend. On the other hand, $A^{SE}$ relates anisotropy to strain energy satisfying the criteria for a clear physical meaning. However, taken in their proposed form, $A^U$, $A^L$ and $A^{SE}$ are strictly applicable to single crystals. 

In addition to anisotropy, certain classes of materials mentioned above such as composite materials, biological tissues, geological materials etc. exhibit heterogeneity. In contrast to homogeneous materials, the elastic stiffness of a heterogeneous material is a function of spatial location of a material point within the material i.e. $C_{ijkl}=C_{ijkl}(x,y,z)$ where $x,y,z \in \mathbb{R}^3$ denote spatial co-ordinates within the material. While the elastic stiffness tensor of a particular single crystal with a distinct crystal structure is unique, the elastic stiffness tensor of a heterogeneous material is a function of the elastic properties and microstructural arrangement of the constituent phases. Consequently, the extent of anisotropy in heterogeneous materials is influenced by the constituent phase properties and microstructural configuration that are crucial descriptors of heterogeneity. However, anisotropy indices proposed in the context of single crystals do not account for the descriptors of heterogeneity thereby limiting their utility in heterogeneous materials. This shortcoming highlights the need to reexamine and reinterpret single crystal anisotropy indices in the context of heterogeneous materials considering the descriptors that influence their overall elastic properties. 

\begin{figure*}[htb]
    \centering
    \begin{subfigure}[t]{0.4\textwidth}
  \includegraphics[width=\textwidth]{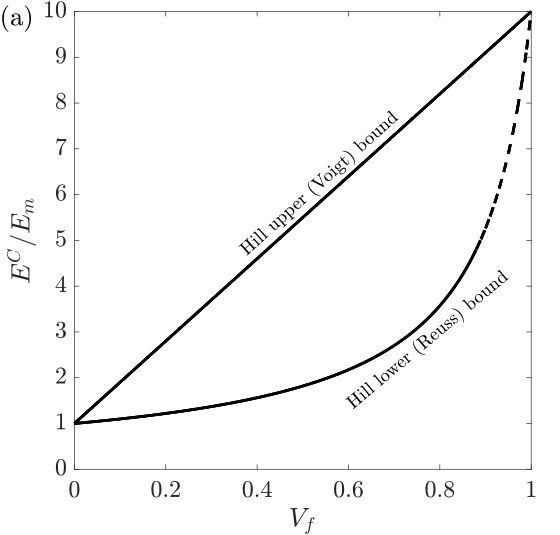}
        \phantomsubcaption
        \label{fig:bounds1}
    \end{subfigure} \qquad
    ~ 
    \begin{subfigure}[t]{0.4\textwidth}      
    \includegraphics[width=\textwidth]{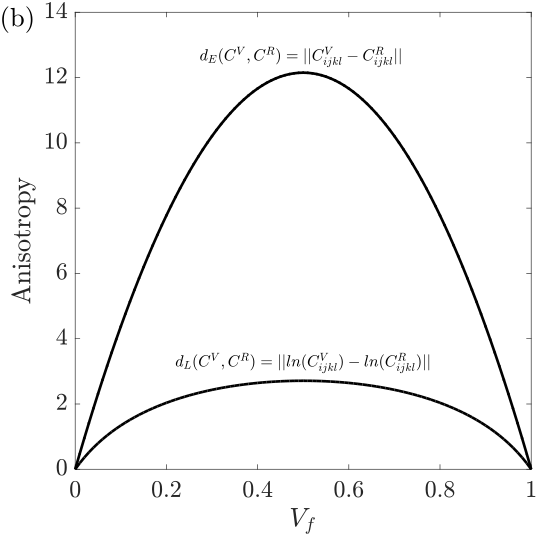}
        \phantomsubcaption
        \label{fig:bounds2}
    \end{subfigure}
    \caption{(a) Hill bounds on the effective elastic properties of 2 phase composite material consisting of isotorpic phases ($E_f/E_m=10$) and (b) Euclidean ($d_E$) and Log-Euclidean ($d_L$) distances between the Hill bounds bounds on the the effective elastic properties of 2 phase composite material  representing the maximum extent of anisotropy.}
\label{fig:Anisotropy_Bounds}
\end{figure*}

This work presents an approach to quantify anisotropy in heterogeneous materials by reinterpreting anisotropy indices proposed for single crystals by considering the effective elastic stiffness tensor of heterogeneous materials in place of the elastic stiffness tensor of single crystals. The effective elastic stiffness tensor evaluated by homogenization embodies the effect of material and microstructural descriptors on the overall elastic response of heterogeneous materials. Therefore, the present approach for anisotropy quantification considering the effective elastic stiffness tensor accounts for the influence of material and microstructural descriptors on the extent of anisotropy. Anisotropy quantification in heterogeneous materials has been demonstrated considering two phase composite materials. Specific material and microstructural descriptors that lead to overall isotropy in two phase composite materials have been rigorously validated. Further, the approach has been extended to quantify anisotropy induced by defects such as pores and cracks in materials thereby demonstrating the generalizability of the approach to any class of heterogeneous materials.

\section{Anisotropy quantification in heterogeneous materials}

The elastic response of a heterogeneous material is dependent on the constituent phase properties, relative content and microstructural configuration. Hill bounds \cite{hill1952} discussed in the case of single crystals while determining distance based anisotropy indices $A^U$ and $A^L$ take a different context in the case of heterogeneous materials \cite{hill1963elastic}. Determination of $A^U$ and $A^L$ proceeds by considering all possible orientations of a single crystal with uniform probability and ensemble averaging leading to isotropic single crystal response. However, heterogeneous materials consist of two or more distinct material phases and Hill bounds considering the individual phase properties provide the absolute bounds on the effective elastic properties \cite{hill1963elastic} as shown in Fig. \ref{fig:bounds1}. The implication is that distance based anisotropy indices $A^U$ and $A^L$ taken in their proposed form (Eq.\ref{eq:anisotropy_ranganathan},\ref{eq:anisotropy_kube}) considering Hill bounds on the effective elastic properties of heterogeneous materials indicate the maximum possible extent of anisotropy as shown in Fig. \ref{fig:bounds2} but not the actual extent of anisotropy. Consequently, anisotropy measures based on the distance between Voigt and Reuss bounds proposed in the context of single crystals need to be suitably modified to be applicable for heterogeneous materials. Strain energy based anisotropy index (Eq.\ref{eq:ASE}) on the other hand requires estimates of the effective elastic stiffness tensor that is a function of the elastic stiffness of the constituent phase and microstructural configuration.

The appropriate approach for anisotropy quantification in a heterogeneous material involves evaluation of its effective elastic properties by homogenization followed by evaluation of anisotropy indices considering the effective elastic stiffness tensor. The effect of material and microstructural descriptors are incorporated in the effective elastic stiffness tensor making this approach amenable to examine their effect on the resulting anisotropy. Details regarding the evaluation of the effective elastic stiffness tensor of a heterogeneous material and anisotropy quantification by reinterpretation of single crystal anisotropy indices have been described in the following section.

\subsection{Effective elastic properties of heterogeneous materials}

\begin{figure*}[htb]
\centering
\includegraphics[width=0.6\textwidth]{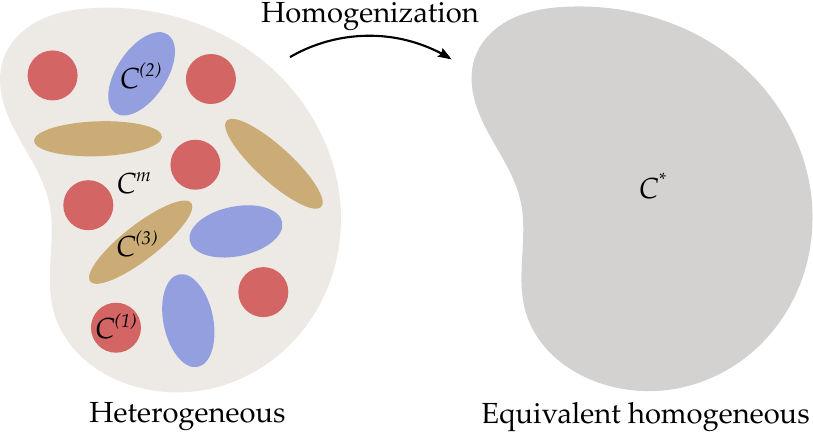}
\caption{Schematic of homogenization representing the evaluation of the effective elastic stiffness $C^*$ of a heterogeneous material consisting of individual phases with elastic stiffness $C^{(1)}$, $C^{(2)}$ etc. embedded in a matrix with elastic stiffness $C^m$.} 
\label{fig:homogenization}
\end{figure*}

The effective elastic stiffness of a heterogeneous material can be obtained by homogenization which is the representation of a heterogeneous material by an equivalent homogeneous material considering individual phase elastic properties and microstructural descriptors (See Fig. \ref{fig:homogenization}). The effective elastic stiffness tensor $C^*_{ijkl}$ of a heterogeneous material can be expressed in terms of volume averaged stress ($\bar{\sigma}_{ij}$) and strain ($\bar{\epsilon}_{kl}$) in the constituent phases of a heterogeneous material in the following form.

\begin{equation}
\bar{\sigma}_{ij}=C^*_{ijkl}\bar{\epsilon}_{kl}
\label{eq:homogenization}
\end{equation}

The effective elastic stiffness $C^*_{ijkl}$ of a heterogeneous material can be obtained from a diverse range of analytical and numerical homogenization approaches \cite{nemat2013micromechanics}. While analytical approaches such as Mori-Tanaka model, composite cylinder assembly, self consistant/generalized self consistant scheme yield closed form solutions for idealized phase geometries, numerical approaches such as method of cells/generalized method of cells, representative volume element/repeating unit cell based numerical homogenization are more versatile in handling complex microstructures.   

The determination of effective elastic stiffness $C^*_{ijkl}$ enables subsequent anisotropy quantification by distance based and strain energy based measures considering $C^*_{ijkl}$ instead of the individual phase properties. Anisotropy quantification in polycrystals \cite{kube2016b} and architectured materials \cite{Wei2021} following a similar approach has been demonstrated previously with the present approach being a generalization that is independent of microstructural configuration and homogenization method and therefore applicable for any class of heterogeneous material.

\subsection{Anisotropy quantification}
The effective elastic stiffness tensor $C^*_{ijkl}$ represents the elastic response of the heterogeneous material under consideration without any restriction on anisotropy. Therefore, the terms in Eq. \ref{eq:anisotropy_ranganathan}, \ref{eq:anisotropy_kube} can be replaced with respective starred terms denoting components of the effective elastic stiffness ($C^*_{ij}$) and compliance ($S^*_{ij}$) matrices in the Voigt representation \cite{hill1952}.

The modified expressions for $A^{U}$ (Eq. \ref{eq:anisotropy_ranganathan}) and $A^{L}$ (Eq. \ref{eq:anisotropy_kube}) are obtained replacing the terms $\kappa^{V}$, $\kappa^{R}$, $\mu^{V}$ and $\mu^{R}$ with respective starred terms $\kappa^{V*}$, $\kappa^{R*}$, $\mu^{V*}$ and $\mu^{R*}$ that reflect their evaluation from $C^*_{ij}$ and $S^*_{ij}$ are as follows.   

\begin{equation}
A^{U*}=\frac{\kappa^{V*}}{\kappa^{R*}}+5 \frac{\mu^{V*}}{\mu^{R*}}-6
\label{eq:au_heterogeneous}  
\end{equation}
\begin{equation}
A^{L*}=\sqrt{\left[ln\left(\frac{\kappa^{V*}}{\kappa^{R*}}\right)\right]^2+5\left[ln \left(\frac{\mu^{V*}}{\mu^{R*}}\right)\right]^2}
\label{eq:al_heterogeneous}  
\end{equation}

Where the terms $\kappa^{V*}$, $\kappa^{R*}$, $\mu^{V*}$ and $\mu^{R*}$ can be expressed in terms of the components of effective elastic stiffness ($C_{ij}^*$) and compliance ($S_{ij}^*$) tensors of a heterogeneous material in Voigt representation in the following form.

\begin{equation}
\label{eq:updatedterms}
\begin{aligned}
9\kappa^{V*} = & \left(C_{11}^*+C_{22}^*+C_{33}^*\right)+2\left(C_{12}^*+C_{13}^*+C_{23}^*\right)\\
15\mu^{V*} = &\left( C_{11}^*+C_{22}^*+C_{33}^*\right)-\left(C_{12}^*+C_{13}^*+C_{23}^* \right)\\
&+3\left(C_{44}^*+C_{55}^*+C_{66}^*\right)\\
1/\kappa^{R*} = &S_{11}^*+S_{22}^*+S_{33}^*+2\left(S_{12}^*+S_{13}^*+S_{23}^*\right)\\
15/\mu^{R*} = &4 (S_{11}^*+S_{22}^*+S_{33}^*-S_{12}^*-S_{13}^*-S_{23}^*)\\
&+3(S_{44}^*+S_{55}^*+S_{66}^*)
\end{aligned}
\end{equation}

A general form of $A^{SE}$  in terms of the strain energy $U$ that is a function of applied strain $\epsilon$ and it's orientation $R$ is given in the following form \cite{fang2019}.

\begin{equation}
A^{SE}=\max_{\epsilon,R^{(1)},R^{(2)}} \left(\frac{U(\epsilon,R^{(1)})}{U(\epsilon,R^{(2)})} \right)-1
\label{eq:ASE_heterogeneous}
\end{equation}

The scalar strain energy $U=\sigma_{ij} \epsilon_{ij}/2$ is a function of the stress and strain tensors involving tensorial operations. Rotation operation in Voigt notation is not consistent since the rotation matrices in stress ($R_{\sigma}$) and strain ($R_{\epsilon}$) spaces differ with both not satisfying the orthogonality relation ($R_{\sigma}R_{\sigma}^T \neq I$,$R_{\epsilon}R_{\epsilon}^T \neq I$). The rotation matrices of strain and stress can be made consistent by using isomorphism between second-rank tensors and six-dimensional vectors as well as between fourth-rank tensors and six-dimensional positive-definite symmetric matrices. The strain and stress tensors can be represented in the isomorphic form ($\tilde{\epsilon}$, $\tilde{\sigma}$) as follows \cite{mehrabadi1995six}.

\begin{equation}
\begin{aligned}
&\tilde{\epsilon}=\begin{Bmatrix}
  \epsilon_{11}\   \epsilon_{22} \ \epsilon_{33}\  \sqrt{2}\epsilon_{23}\   \sqrt{2}\epsilon_{13}\   \sqrt{2}\epsilon_{12} \ 
 \end{Bmatrix}^T\\
&\tilde{\sigma}=\begin{Bmatrix}
  \sigma_{11}  \ \sigma_{22} \ \sigma_{33}  \ \sqrt{2}\sigma_{23}  \ \sqrt{2}\sigma_{13}  \ \sqrt{2}\sigma_{12}
 \end{Bmatrix}^T
 \end{aligned}
  \label{eq:isomorphic}  
\end{equation}

The elastic stiffness tensor in Voigt notation ($C$) represented in the isomorphic form $\tilde{C}$ is as follows \cite{mehrabadi1995six}.

\begin{equation}
  \tilde{C}=
 \begin{bmatrix}
  C_{11} & C_{12} & C_{13} & \sqrt{2}C_{14} & \sqrt{2}C_{15} & \sqrt{2}C_{16} \\
   & C_{22} & C_{23} & \sqrt{2}C_{24} & \sqrt{2}C_{25} & \sqrt{2}C_{26} \\  
   &  & C_{33} & \sqrt{2}C_{34} & \sqrt{2}C_{35} & \sqrt{2}C_{36} \\
   &  &  & 2C_{44} & 2C_{45} & 2C_{46} \\ 
   & sym &  &  & 2C_{55} & 2C_{56} \\
   &  &  &  &  & 2C_{66} \\     
 \end{bmatrix}
  \label{eq:constitutive3}  
\end{equation}

The isomorphic representation of the terms in the generalized Hooke's law retains it's representation i.e. $\tilde{\sigma}=\tilde{C}\tilde{\epsilon}$ as well as the  expression for elastic strain energy in the following form  

\begin{equation}
\begin{aligned}
& U= \frac{1}{2} \epsilon^T . C . \epsilon = \frac{1}{2} \tilde{\epsilon}^T . \tilde{C} . \tilde{\epsilon} \\
 \end{aligned}
  \label{eq:strainenergy}  
\end{equation}

 Rotation of $\tilde{\epsilon}$ , $\tilde{\sigma}$ and $\tilde{C}$ can be accomplished by the modified rotation matrix $\tilde{R}$ shown in Eq. \ref{eq:rotationmatrix} \cite{mehrabadi1995six,norris2008euler} that consists of directional cosines $Q_{ij}=cos(x_i',x_j)$ $i,j=1,2,3$. The modified rotation matrix $\tilde{R}$ satisfies the orthoganality relation i.e. $\tilde{R}\tilde{R}^T=\tilde{R}^T\tilde{R}=I$ leading to the rotation matrices of strain and stress being consistent.

\begin{widetext}
 \begin{equation}
   \tilde{R} = 
  \begin{bmatrix}
   Q_{11}^2& Q_{12}^2& Q_{13}^2& \sqrt{2}Q_{12}Q_{13}& \sqrt{2}Q_{11}Q_{13}& \sqrt{2} Q_{11}Q_{12} \\
   Q_{21}^2& Q_{22}^2& Q_{23}^2& \sqrt{2}Q_{22}Q_{23}& \sqrt{2}Q_{21}Q_{23}& \sqrt{2}Q_{21}Q_{22} \\  
   Q_{31}^2& Q_{32}^2& Q_{33}^2& \sqrt{2} Q_{32} Q_{33}& \sqrt{2} Q_{31} Q_{33}& \sqrt{2} Q_{31} Q_{32} \\
   \sqrt{2} Q_{21} Q_{31}& \sqrt{2} Q_{22} Q_{32}& \sqrt{2} Q_{23} Q_{33}& Q_{22} Q_{33} + Q_{23} Q_{32}& Q_{21} Q_{33} + Q_{23} Q_{31}& Q_{21} Q_{32} + Q_{22} Q_{31}  \\ 
   \sqrt{2} Q_{11} Q_{31}& \sqrt{2} Q_{12} Q_{32}& \sqrt{2} Q_{13} Q_{33}& Q_{12} Q_{33} + Q_{13} Q_{32}& Q_{11} Q_{33} + Q_{13} Q_{31}& Q_{11} Q_{32} + Q_{12} Q_{31} \\
   \sqrt{2} Q_{11} Q_{21}& \sqrt{2} Q_{12} Q_{22}& \sqrt{2} Q_{13} Q_{23}& Q_{12} Q_{23} + Q_{13} Q_{22}& Q_{11} Q_{23} + Q_{13} Q_{21}& Q_{11} Q_{22} + Q_{12} Q_{21} \\     
  \end{bmatrix}
   \label{eq:rotationmatrix}
 \end{equation}
\end{widetext}

Utilizing the isomorphic representation of generalized Hooke's law and expressions for coordinate transformation, the strain energy based anisotropy index in strain space for heterogeneous materials is given by the following expression

\begin{equation}
A^{SE*}=\max_{\epsilon,\tilde{R}^{(1)},\tilde{R}^{(2)}}\left(\frac{(\tilde{R}^{(1)}\epsilon)^T \tilde{C^*} (\tilde{R}^{(1)}\epsilon)}{(\tilde{R}^{(2)}\epsilon)^T \tilde{C^*} (\tilde{R}^{(2)}\epsilon)} \right)-1 
\label{eq:ASE2}
\end{equation}

where $\tilde{R}^{(1)}$ and $\tilde{R}^{(2)}$ represents the rotation matrices for  distinct orientations and $\epsilon$ represents applied strain.

\subsection{Numerical implementation and validation}

Distance based anisotropy indices $A^{U}$ and $A^{L}$ have closed form expressions given by Eq.\ref{eq:anisotropy_ranganathan} and Eq.\ref{eq:anisotropy_kube} respectively that can be directly evaluated from components of stiffness and compliance tensors. However, evaluation of strain energy based anisotropy index $A^{SE}$ requires optimization through appropriate optimization algorithms. In the present work, evaluation of $\epsilon$,  $\tilde{R}^{(1)}$ and $\tilde{R}^{(2)}$ that maximize the function in Eq.\ref{eq:ASE} and Eq.\ref{eq:ASE2} has been performed by genetic algorithm in MATLAB (R2017a). Initial population size of 1000 and function tolerance of 10\textsuperscript{-3} have been used for optimization with the genetic algorithm. The numerical implementation of the approaches to evaluate anisotropy indices $A^U$,$A^L$,$A^{SE}$ can be validated by the following means. 

(a) fulfillment of the second criteria for an acceptable anisotropy index i.e. anisotropy index tends to infinity ($A\rightarrow\infty$) if there exists a direction along which the deformation resistance of the material approaches zero ($C_{ij}\rightarrow 0$). This can be evaluated by considering a hypothetical stiffness matrix in the following form \cite{fang2019}.

\begin{equation}
  C_{ij}=
 \begin{bmatrix}
  2\times10^{n-1} & 1 & 1 & 0 & 0 & 0 \\
   & 2 & 1 & 0 & 0 & 0 \\  
   &  & 2 & 0 & 0 & 0 \\
   &  &  & 1 & 0 & 0 \\ 
   & sym &  &  & 1 & 0 \\
   &  &  &  &  & 1 \\     
 \end{bmatrix};\ n=1,2,3,...
  \label{eq:conditionnumber}  
\end{equation}

\begin{figure*}[htb]
    \centering
    \begin{subfigure}[t]{0.4\textwidth}
  \includegraphics[width=\textwidth]{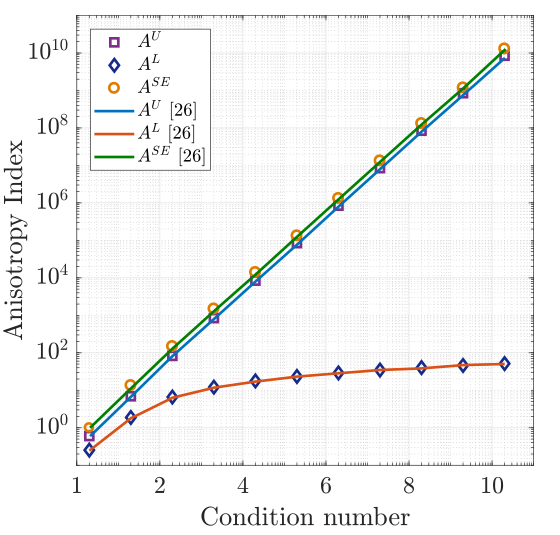}
        \phantomsubcaption
        \label{fig:validation1}
    \end{subfigure} 
    ~ 
    \begin{subfigure}[t]{0.4\textwidth}      
    \includegraphics[width=\textwidth]{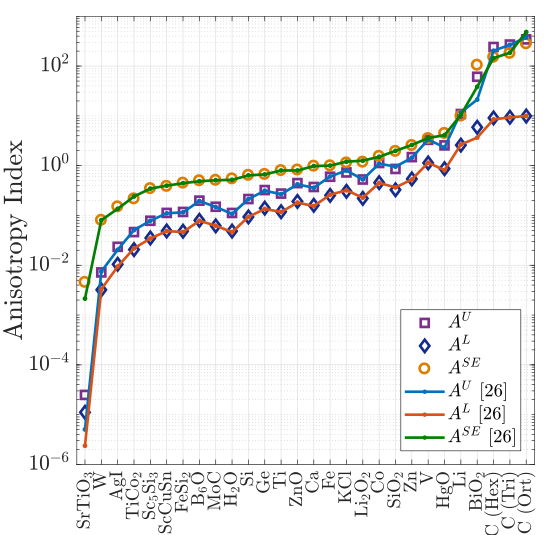}
        \phantomsubcaption
        \label{fig:validation2}
    \end{subfigure}
    \vspace{-10pt}
    \caption{Validation of the approach to evaluate the Anisotropy indices ($A^U$,$A^L$,$A^{SE}$) of single crystals (a) effect of condition number on the anisotropy indices and (b) prediction of anisotropy indices of selected single crystals. Lines represent results from \cite{fang2019} and symbols represent results from the present implementation.}
    \label{fig:validation3}
\end{figure*}

As the `condition number' $n$ increases, the extent of anisotropy increases that should reflect in the anisotropy indices. Anisotropy indices $A^U$,$A^L$,$A^{SE}$ evaluated in the present work compared with the values reported in \cite{fang2019} has been presented in Fig. \ref{fig:validation1}. A good agreement in the values validate the numerical implementation of anisotropy index evaluation.

(b) comparing the predictions from current implementation with the anisotropy indices presented for selected single crystals in \cite{fang2019}. Prediction of anisotropy indices by the present implementation across a set of single crystals (accessed at `The Materials Project' \cite{materialsproject}) has been presented in Fig. \ref{fig:validation2} that shows a good agreement with the results presented in \cite{fang2019} except for deviations for BiO\textsubscript{2}, validating the procedure to evaluate $A^U$, $A^L$ and $A^{SE}$.

\section{Results and Discussion}

\subsection{Anisotropy in particulate composites}

Anisotropy quantification in heterogeneous materials can demonstrated by considering two phase composite materials due to the availability of closed form expressions for idealized reinforcement shapes. Considering a two phase composite material consisting of aligned isotropic ellipsoidal particulates embedded in isotropic matrix, the effective elastic stiffness tensor accounting for the elastic properties and  volume fractions of the constituent phases along with the shape of the particulates is given by Mori-Tanaka approach \cite{benveniste1987new} in the following form.

\begin{equation}
C^{*}=(1-V_f ) C^m  A^{MT}+V_f C^f  A^{MT} A^{Eshelby}
\end{equation}

\noindent where $V_f$ is the volume fraction of the particulate phase, $C^f$ and $C^m$ are the stiffness tensors of the particulate and matrix phase respectively and $A^{MT}$ is the Mori-Tanaka strain concentration tensor given by 

\begin{equation}
A^{MT}=[(1-V_f )I+V_f A^{Eshelby} ]^{-1}
\label{eq:AMT}
\end{equation}

\noindent where $I$ is identity matrix and $A^{Eshelby}$ is the Eshelby strain concentration tensor in terms of the Eshelby tensor $S$ \cite{eshelby1957determination} that accounts for the particulate shape.

\begin{equation}
A^{Eshelby}=[I+S C^m (C^f-C^m )]^{-1}
\label{eq:AESHELBY}
\end{equation}

The advantage of Mori-Tanaka approach is the ability to consider shape parameters of general ellipsoidal shapes along with volume fractions and constituent phase properties in effective property estimation. With the determination of $C^{*}$, extent of anisotropy in two phase composite materials can be evaluated and the influence of constituent properties, volume fractions, particulate shape and aspect ratio on anisotropy can be assessed. 

\begin{figure*}[htb]
    \centering
    \begin{subfigure}[t]{0.4\textwidth}
    \includegraphics[width=\textwidth]{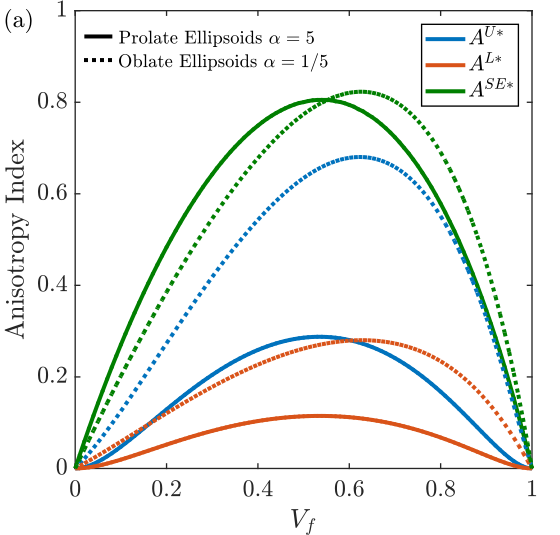}
    \phantomsubcaption
    \label{fig:anisotropy1}
    \end{subfigure} \qquad 
    \begin{subfigure}[t]{0.4\textwidth}      
    \includegraphics[width=\textwidth]{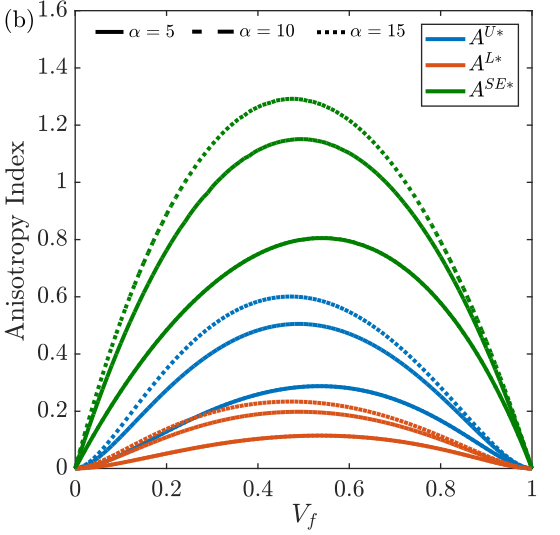}
    \phantomsubcaption
    \label{fig:anisotropy2}
    \end{subfigure}\\
    \begin{subfigure}[t]{0.4\textwidth}      
    \includegraphics[width=\textwidth]{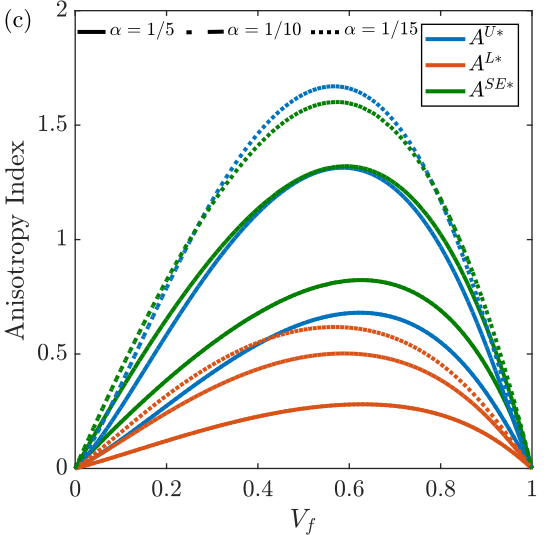}
    \phantomsubcaption
    \label{fig:anisotropy3}
    \end{subfigure} \qquad
    \begin{subfigure}[t]{0.4\textwidth}      
    \includegraphics[width=\textwidth]{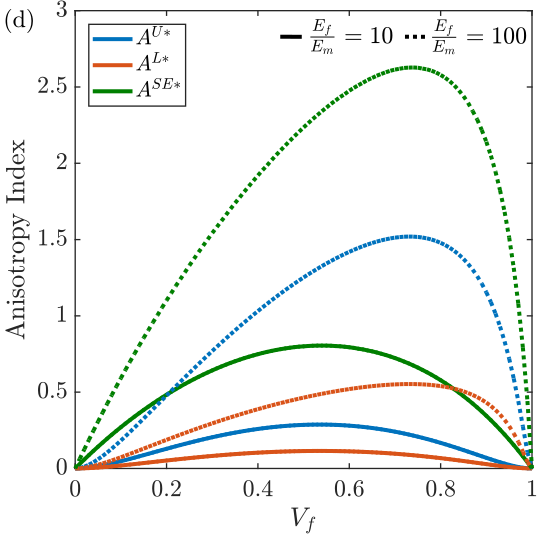}
    \phantomsubcaption
    \label{fig:anisotropy4}
    \end{subfigure}
    \vspace{-10pt}           
    \caption{Influence of microstructural and material descriptors on anisotropy in two phase particulate reinforced composite materials: effect of (a) overall shape of the particulates and particulate volume fraction ($E_f/E_m=10$), (b) aspect ratio of prolate ellipsoids and (c) aspect ratio of prolate ellipsoids both highlighting the influence of microstructural descriptors ($E_f/E_m=10$), and (d) elastic contrast ratio highlighting the influence of material descriptors (prolate ellipsoids, $\alpha=5$).}
    \label{fig:anisotropy_1}
\end{figure*}

The microstructural features that influence $C^*$ and consequently the extent of anisotropy in two phase particle reinforced composite materials are the reinforcement volume fraction, reinforcement shape and aspect ratio of a particular shape under consideration. Assuming that the reinforcement and matrix are individually isotropic with elastic contrast ratio $E_f/E_m=10$, the role of volume fractions and overall shape of the particulates in influencing anisotropy can be inferred from the variation of anisotropy indices with particulate volume fractions as shown in Fig. \ref{fig:anisotropy1}. The anisotropy indices increase from zero, reach a peak and reduce back to zero. The extremities of the volume fraction range i.e. $V_f=0,1$ have anisotropy indices equal to zero since they denote a single material phase, either matrix ($V_f=0$) or reinforcements ($V_f=1$) that are individually isotropic. Deviation from isotropy in the intermediate particulate volume fraction range ($0<V_f<1$) implies that the presence of a secondary phase with elastic property mismatch causes anisotropy even though the constituents are individually isotropic. 

The nature of variation of anisotropy indices differ across prolate and oblate ellipsoidal particulates (Fig. \ref{fig:anisotropy1}) highlighting the effect of overall shape of secondary phase on the extent of anisotropy. Further, anisotropy increases with increasing aspect ratio ($\alpha$) across the range of volume fractions in the case of prolate ellipsoids (Fig. \ref{fig:anisotropy2}) and oblate ellipsoids (Fig. \ref{fig:anisotropy3}) highlighting the effect of shape parameters on anisotropy. The overall shape and the shape parameters along with the particulate volume fractions can be collectively termed as the microstructural descriptors that influence anisotropy in two phase composite materials.

\begin{figure*}[htb]
    \centering
    \begin{subfigure}[t]{0.4\textwidth}
    \includegraphics[width=\textwidth,trim= {0cm 0.5cm 0cm 0.5cmcm},clip]{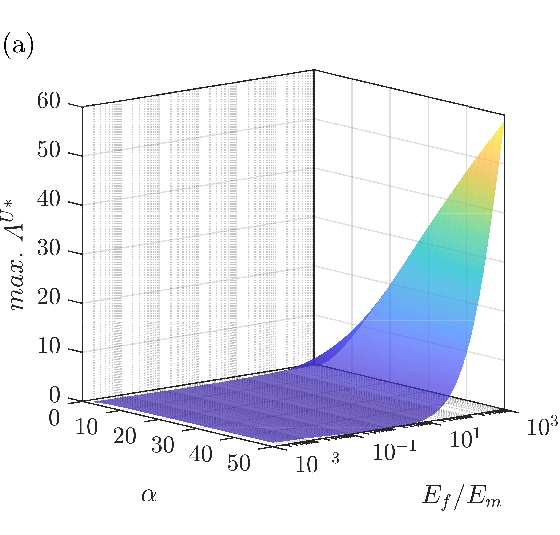}
    \phantomsubcaption
    \label{fig:anisotropy_map1}        
    \end{subfigure}
    \qquad
    \begin{subfigure}[t]{0.4\textwidth}
    \includegraphics[width=\textwidth,trim= {0cm 0.5cm 0cm 0.5cmcm},clip]{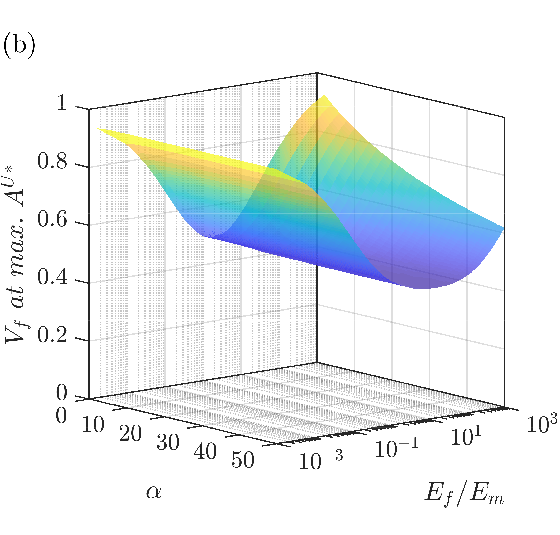}
    \phantomsubcaption
    \label{fig:anisotropy_map2}
    \end{subfigure}
    \begin{subfigure}[t]{0.4\textwidth}
    \includegraphics[width=\textwidth,trim= {0cm 0.5cm 0cm 0.5cmcm},clip]{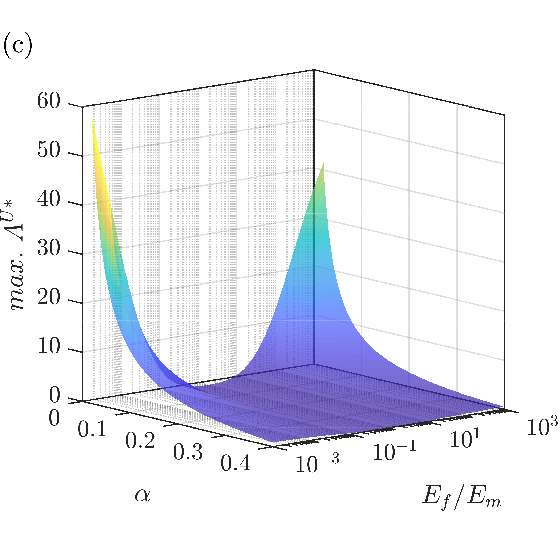}
    \phantomsubcaption
    \label{fig:anisotropy_map3}        
    \end{subfigure}
    \qquad
    \begin{subfigure}[t]{0.4\textwidth}
    \includegraphics[width=\textwidth,trim= {0cm 0.5cm 0cm 0.5cmcm},clip]{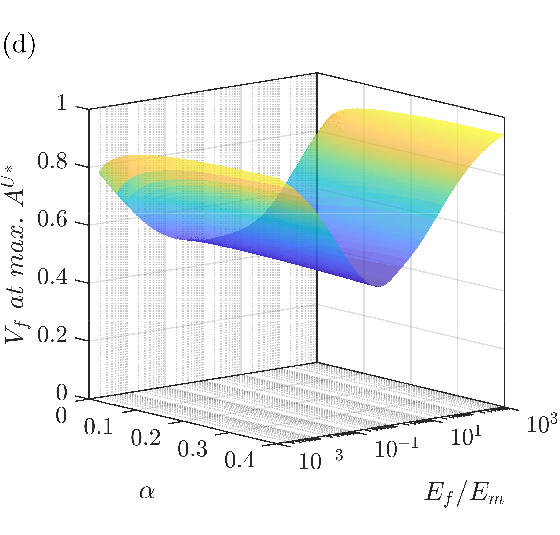}
    \phantomsubcaption
    \label{fig:anisotropy_map4}
    \end{subfigure}  
    \vspace{-10pt}      
    \caption{Peak anisotropy map for ellipsoidal particle reinforced composites representing the variation of maximum anisotropy index ${A^{U*}}$ and the corresponding volume fraction with aspect ratio ($\alpha$) and elastic contrast ratio ($E_F/E_m$): (a) variation of $\max.\ {A^{U*}}$ for prolate ellipsoids, (b) $V_f$ at $\max.\ {A^{U*}}$ for prolate ellipsoids, (c) variation of $\max.\ {A^{U*}}$ for oblate ellipsoids, and (d) $V_f$ at $\max.\ {A^{U*}}$ for oblate ellipsoids.}
    \label{fig:anisotropy_2}
\end{figure*}

Elastic contrast, defined as the ratio of the Young's moduli of the constituent phases ($E_f/E_m$), is a crucial material descriptor in composite materials. Considering prolate ellipsoidal particulates, elastic contrast has a significant influence on the extent of anisotropy across the range of volume fractions as seen in Fig. \ref{fig:anisotropy4}. Variation in elastic contrast not only changes the extent of anisotropy across the range of volume fractions but also influences the peak anisotropy and the particulate volume fraction at which the peak anisotropy occurs. The significant influence of elastic contrast on the extent of anisotropy makes it a key material descriptor influencing anisotropy in two phase composite materials in addition to microstructural descriptors discussed previously.

An important aspect of heterogeneity driven anisotropy is the maximum  anisotropy representing the maximum deviation from isotropy due to microstructural and material descriptors. The influence of microstructural and material descriptors on the maximum anisotropy can be inferred from a plot of the variation of maximum elastic anisotropy in two phase composite microstructures with isotropic constituent phases as shown in Fig. \ref{fig:anisotropy_2}. Prolate (Fig. \ref{fig:anisotropy_map1}) and oblate (Fig. \ref{fig:anisotropy_map3}) ellipsoidal reinforcement exhibit vastly differing trends in the variation of maximum anisotropy signifying the influence of overall shape of reinforcement. High aspect ratio  stiffer reinforcements ($E_f/E_m$ >> 1) lead to maximum anisotropy being higher in the case of prolate ellipsoids. In contrast, high elastic ratio mismatch irrespective of stiffer ($E_f/E_m$ >> 1) or compliant ($E_f/E_m$ << 1) reinforcement lead to maximum anisotropy being higher in the case of oblate ellipsoids.

The reinforcement volume fraction corresponding to maximum anisotropy provides insights into the relative content of constituent phases leading to maximum deviation from isotropy. Plots of maximum anisotropy and reinforcement volume fraction corresponding to maximum anisotropy have been presented for prolate (Fig. \ref{fig:anisotropy_map2}) and oblate (Fig. \ref{fig:anisotropy_map4}) ellipsoidal reinforcements that exhibit differing trends indicating the influence of reinforcement shape. 

\begin{figure*}[htb]
    \centering
    \begin{subfigure}[t]{0.4\textwidth}
    \includegraphics[width=\textwidth]{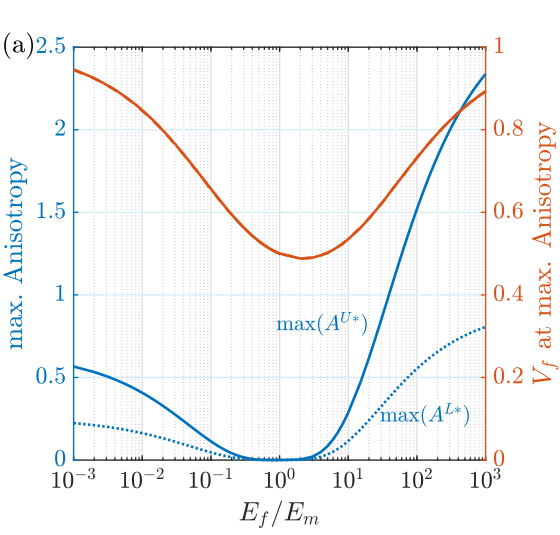}
    \phantomsubcaption
    \label{fig:max_anisotropy_map1}        
    \end{subfigure}
    \qquad
    \begin{subfigure}[t]{0.4\textwidth}
    \includegraphics[width=\textwidth]{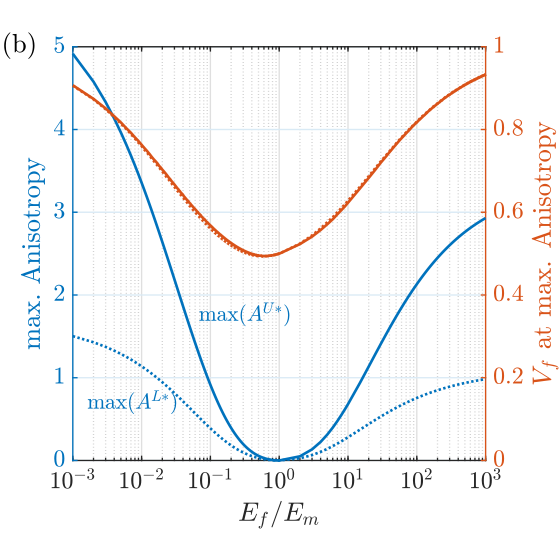}
    \phantomsubcaption
    \label{fig:max_anisotropy_map2}
    \end{subfigure}  
    \vspace{-10pt}      
    \caption{Maximum anisotropy and particulate volume fraction at which maximum anisotropy occurs for ellipsoidal particulate reinforced composite material with aligned ellipsoidal reinforcements: (a) prolate ellipsoids, $\alpha=5$, and (b) oblate ellipsoids, $\alpha=1/5$.}
    \label{fig:anisotropy_3}
\end{figure*}

The variation of maximum anisotropy and volume fraction corresponding to maximum anisotropy with prolate and oblate ellipsoidal reinforcement of particular aspect ratio has been presented in Fig. \ref{fig:anisotropy_3} to highlight the effect of elastic contrast. While $A^{U*}$ and $A^{L*}$ exhibit differing magnitudes of maximum anisotropy as expected, the volume fraction corresponding to maximum anisotropy evaluated by the two indices coincide.

The most distinct trend indicated by the variation of reinforcement volume fraction corresponding to maximum anisotropy is that maximum anisotropy consistently occurs above reinforcement volume fraction greater than 40\% across the range of contrast ratios in both prolate and oblate ellipsoidal reinforcements. Elastic anisotropy peaking at the higher end of the volume fraction range indicates the existence of a threshold volume fraction corresponding to maximum anisotropy that depends on the microstructural and material descriptors. In the case of engineering composites ($0.1 < E_f/E_m < 100$),  maximum anisotropy occurs within a much narrow reinforcement volume fraction range i.e. between 45\% -60\% that overlaps with the typical range of reinforcement volume fractions.

\subsection{Overall isotropy in particulate composites}

\begin{figure*}[htb]
    \begin{subfigure}[t]{0.4\textwidth}
    \includegraphics[width=\textwidth,trim= {0.0cm 4.5cm 1.0cm 5.6cm},clip]{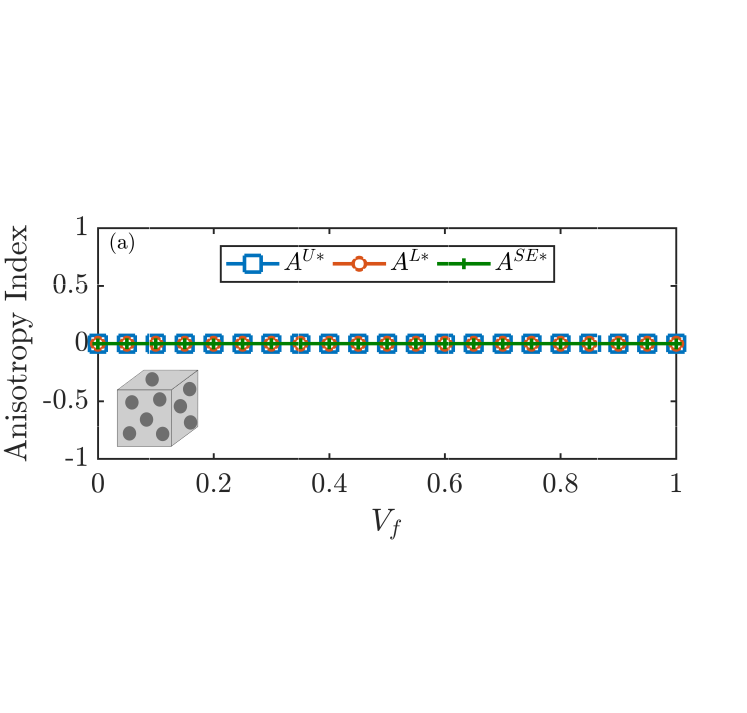}
    \phantomsubcaption
    \label{fig:spheres}
    \end{subfigure} \qquad
    \begin{subfigure}[t]{0.4\textwidth}      
    \includegraphics[width=\textwidth,trim= {0.0cm 4.5cm 1.0cm 5.6cm},clip]{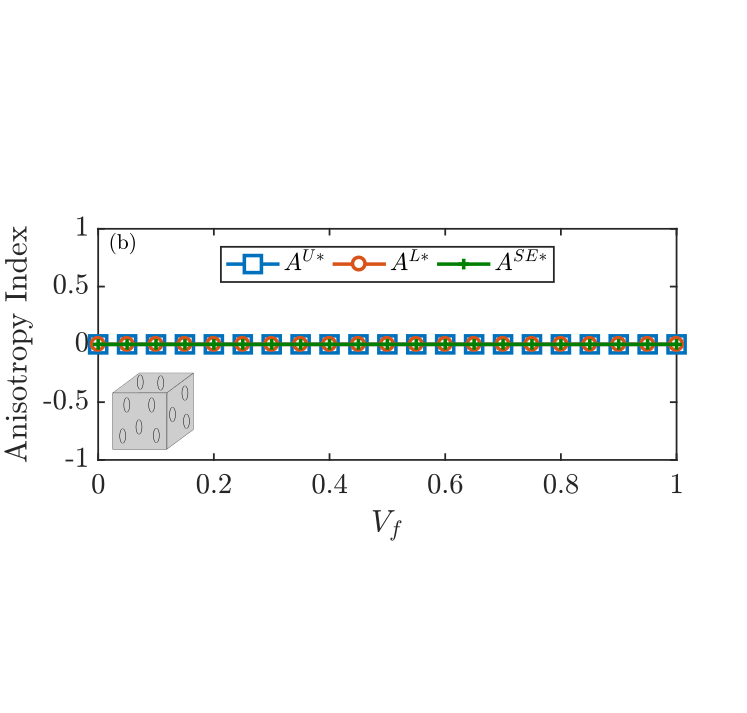}
    \phantomsubcaption
    \label{fig:ContrastRatio1}
    \end{subfigure} \\
    \begin{subfigure}[t]{0.4\textwidth}      
    \includegraphics[width=\textwidth,trim= {0.0cm 4.5cm 1.0cm 5.6cm},clip]{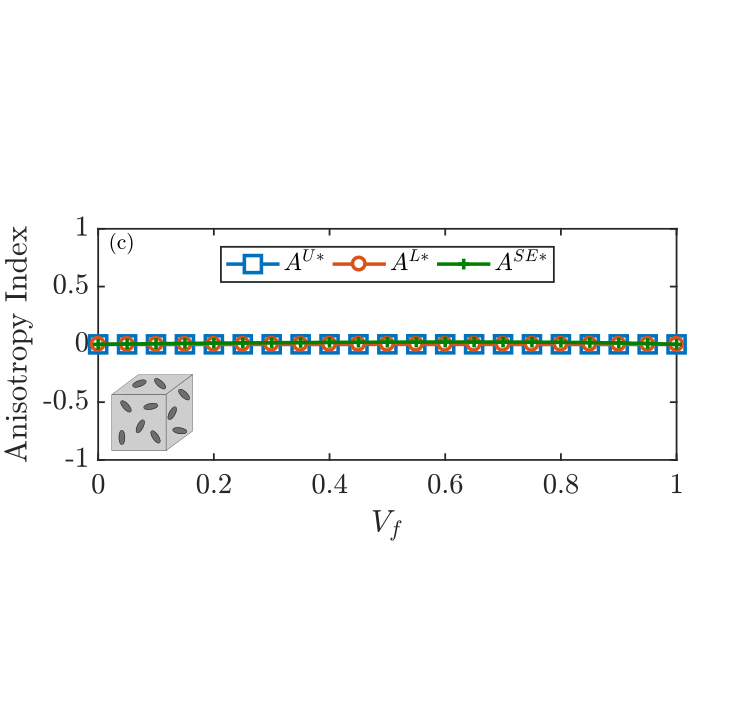}
    \phantomsubcaption
    \label{fig:RandomProlate}
    \end{subfigure} \qquad
    \begin{subfigure}[t]{0.4\textwidth}      
    \includegraphics[width=\textwidth,trim= {0.0cm 4.5cm 1.0cm 5.6cm},clip]{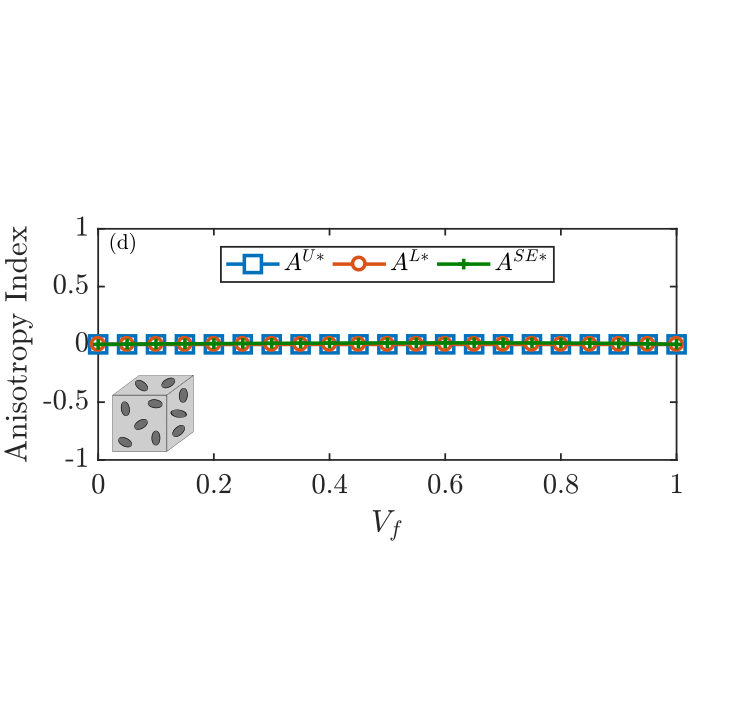}
    \phantomsubcaption
    \label{fig:RandomOblate}
    \end{subfigure}         
    \caption{Two phase composite materials consisting of isotropic ellipsoidal particulates in isotropic matrix lead to overall isotropy indicated by anisotropy indices $A^U$,$A^L$,$A^{SE}$ being zero under the following conditions (composite configuration mentioned in brackets): (a) particulates are spherical ($E_f/E_m=10$), (b) aligned particulates of general ellipsoidal shapes have a contrast ratio of 1 (aligned prolate ellipsoids, $\alpha=5$, $E_f/E_m=1$), (c) randomly oriented prolate ellipsoids ($\alpha=5$, $E_f/E_m=10$), and (d) randomly oriented oblate ellipsoids ($\alpha=1/5$, $E_f/E_m=10$).}
    \label{fig:anisotropy_4}
\end{figure*}

In the backdrop of heterogeneity driven anisotropy, the conditions that lead to overall isotropy in heterogeneous material systems is of special interest. Overall isotropy can be inferred by anisotropy indices $A^{U*}$, $A^{L*}$ and $A^{SE*}$ being consistently zero with a variation in material and microstructural descriptors. Two phase particle reinforced composite materials exhibit overall isotropy indicated by anisotropy indices going to zero when the particulates are spherical (Fig.\ref{fig:spheres}) and when the contrast ratio is unity i.e. $E_f/E_m=1$ (Fig.\ref{fig:ContrastRatio1}). Spheres have identical response in all directions leading to overall isotropy in heterogeneous materials with spherical particulates. Contrast ratio being unity signifies identical properties of the constituent phases that is analogous to homogeneous material.

While isotropic matrices consisting of randomly oriented reinforcement phases of ellipsoidal shapes are stated to exhibit overall isotropy, the present approach enables a rigorous proof of the same. Consider a two phase particle reinforced composite material with randomly oriented isotropic ellipsoidal particles embedded in isotropic matrix. Random distribution of the reinforcing phases can be modeled by co-ordinate transformation of the strain concentration tensor and averaging the transformed strain concentration tensor over all possible orientations. Strain concentration tensor follows the transformation rule for fourth order tensors i.e. 

\begin{equation}
\label{eq:Rot}
\bar{A}^{Eshelby}_{ijkl}=Q_{ip} Q_{jp} Q_{kr} Q_{ls} A^{Eshelby}_{pqrs}
\end{equation}

where $Q_{ij}$ represents the rotation tensor given by $Q_{ij}=cos(x_i',x_j)$ $i,j=1,2,3$. Averaging the strain concentration tensor over the range of orientations leads to the overall response with randomly distributed reinforcement phase. Orientation averaged Eshelby strain concentration tensor, where orientation averaging is denoted by '$\left\langle \right\rangle$', can be represented in the following form \cite{ferrari1989}. 

\begin{equation}
\label{eq:Arot}
\left\langle{A^{Eshelby}}\right\rangle = \dfrac{\int_{-\pi}^{\pi}\int_{0}^{\pi}\int_{0}^{\pi/2}[\bar{A}^{Eshelby}(\phi,\gamma,\psi) \ sin\gamma \ d\phi\ d\gamma\ d\psi]}{\int_{-\pi}^{\pi}\int_{0}^{\pi}\int_{0}^{\pi/2} sin\gamma \ d\phi\ d\gamma\ d\psi}
\end{equation}

Where $\gamma,\phi,\psi$ represent orientations w.r.t global co-ordinates. The Mori-Tanaka strain concentration tensor for random orientation of particulate phases is given by the following expression.

\begin{equation}
A^{MT}_{Random}=[(1-V_f )I+V_f \left\langle{A^{Eshelby}}\right\rangle ]^{-1}
\label{eq:MTrot}
\end{equation}

The effective elastic properties with randomly oriented particulate phase ($C^{*}_{Random}$) can be evaluated considering the orientation averaged strain concentration tensor in the following form.

\begin{equation}
 C^{*}_{Random}=(1-V_f ) C^m  A^{MT}_{Random}+V_f C^f  A^{MT}_{Random} \left\langle A^{Eshelby}\right\rangle
\label{eq:moritanakaaveraged}
\end{equation}

Evaluation of $A^U$, $A^L$ and $A^{SE}$ considering $C^{*}_{Random}$ corresponding to randomly oriented ellipsoidal particulates reveal that they are zero across the range of volume fractions (Figs.\ref{fig:RandomProlate},\ref{fig:RandomOblate}) indicating overall isotropy irrespective of overall shape, shape parameters and contrast ratio. Previous observations regarding overall isotropy in 2 phase composites with randomly oriented ellipsoidal particles (Benveniste \cite{benveniste1987new} and references therein) can be rigorously proved with the present approach. However, such observations regarding overall isotropy cannot be extended directly to the case where the constituents themselves are anisotropic.

\subsection{Defect driven elastic anisotropy}

Homogeneous elastic medium containing dry pores and cracks can be considered as a special case of heterogeneous materials with the pores/cracks representing a distinct phase with zero elastic stiffness \cite{zhao1989elastic,pan1995elastic}.
The effective elastic stiffness tensor of a material consisting of ellipsoidal pores $C^{*}_{Pore}$ accounting for the pore volume fraction $V_P$ and shape of pores is given by the Mori-Tanaka approach in the following form.

\begin{equation}
C^{*}_{Pore}=(1-V_P ) C^m  A^{MT}
\end{equation}

\noindent where $C^m$ is the stiffness tensor of the matrix phase and $V_P$ is the pore volume fraction. Mori-Tanaka strain concentration tensor $A^{MT}$ and Eshelby strain concentration $A^{Eshelby}$ are obtained for the case of pores by replacing $V_f$ with $V_P$ and $C^f$ by a null matrix in Eq. \ref{eq:AMT} and Eq. \ref{eq:AESHELBY} respectively. Evaluation of anisotropy indices with the effective elastic stiffness of  homogeneous material containing ellipsoidal pores ($C^{*}_{Pore}$) following the approach for anisotropy quantification in two phase composite materials provides an insight into the extent of anisotropy caused by the presence of porosity in homogeneous materials.

\begin{figure*}[tb]
    \centering
    \begin{subfigure}[t]{0.4\textwidth}
    \includegraphics[width=\textwidth]{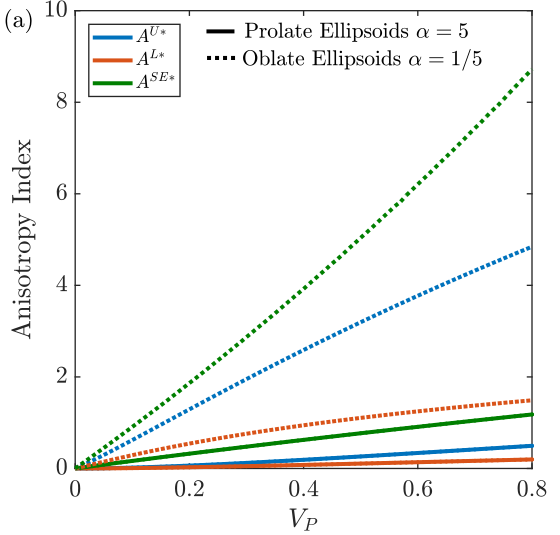}
    \phantomsubcaption
    \label{fig:anisotropy_pores1}        
    \end{subfigure} \qquad
    \begin{subfigure}[t]{0.4\textwidth}
    \includegraphics[width=\textwidth]{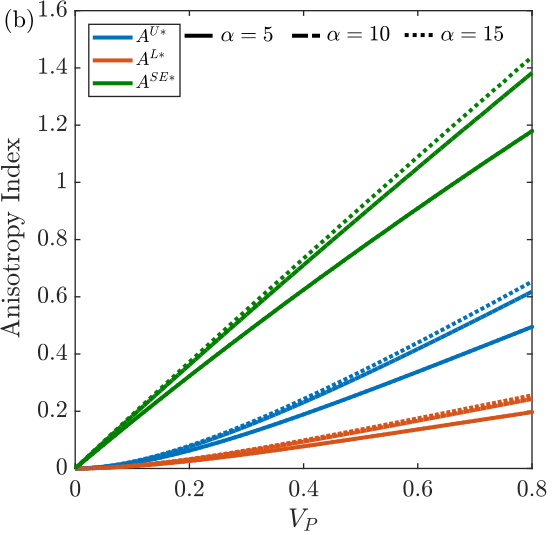}
    \phantomsubcaption
    \label{fig:anisotropy_pores2}
    \end{subfigure} \\ 
    \begin{subfigure}[t]{0.4\textwidth}      
    \includegraphics[width=\textwidth]{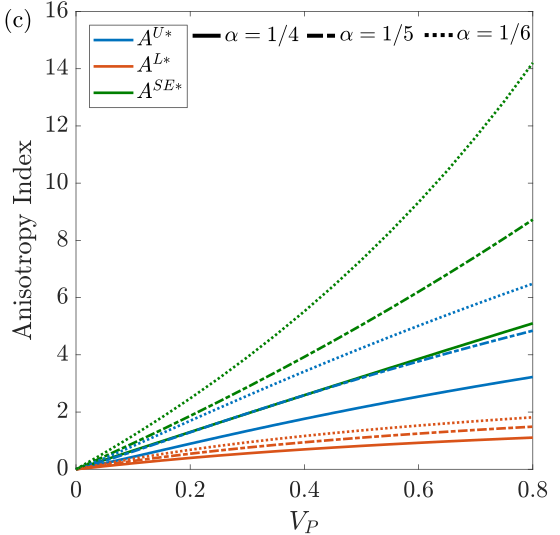}
    \phantomsubcaption
    \label{fig:anisotropy_pores3}
    \end{subfigure} \qquad
    \begin{subfigure}[t]{0.4\textwidth}      
    \includegraphics[width=\textwidth]{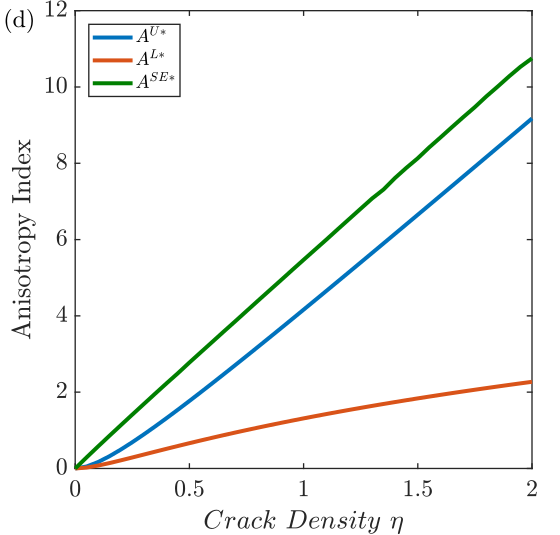}
    \phantomsubcaption
    \label{fig:anisotropy_cracks}
    \end{subfigure}
    \vspace{-10pt}      
    \caption{Influence of pores and cracks on anisotropy in isotropic homogeneous materials: (a) variation of anisotropy with pore volume fraction $V_p$ highlighting the effect of overall shape of aligned ellipsoidal pores on anisotropy, (b) effect of aspect ratio of aligned prolate ellipsoidal pores emphasizing the influence of shape parameters, (c) variation of anisotropy with crack density $\eta$, and (d) overall isotropy in isotropic media consisting of randomly oriented prolate ellipsoidal pores ($\alpha=2$) or cracks indicated by anisotropy indices $A^U$,$A^L$,$A^{SE}$ being zero.}
    \label{fig:anisotropy_5}
\end{figure*}

The extent of anisotropy brought about by the presence of aligned ellipsoidal pores in an isotropic homogeneous material across the range of pore volume fraction $V_P$ has been shown in Fig. \ref{fig:anisotropy_pores1}. Anisotropy in porous materials increases monotonically with pore volume fraction ($V_P$) in stark contrast to two phase heterogeneous materials. Pore shape has a significant effect on the nature of anisotropy variation with prolate and oblate ellipsoidal pores exhibiting markedly different trends as shown in Fig. \ref{fig:anisotropy_pores1}. Across the pore volume fractions, increasing pore aspect ratios leads to increase in anisotropy in prolate ellipsoidal pores (Fig. \ref{fig:anisotropy_pores2})  and decrease in anisotropy in oblate ellipsoidal pores (Fig. \ref{fig:anisotropy_pores3}). Thus, the pore volume fraction, pore shape and aspect ratio form the microstructural descriptors that influence anisotropy in porous materials.

Circular cracks represent the extreme case of collapsed discus shaped pores with the aspect ratio tending to zero ($\alpha\rightarrow0$). Consequently,  the volume of individual cracks tend to zero requiring quantification of crack content by crack density `$\eta$' that is related to the crack volume fraction $V_C$ in the following form.

\begin{equation}
\eta=\frac{3 V_C}{4\pi\alpha}
\end{equation}

\noindent where $\alpha$ represents the aspect ratio. Considering homogeneous isotropic medium containing aligned circular cracks, the variation of anisotropy indices with crack density has been presented in Fig. \ref{fig:anisotropy_cracks}. It can be seen that the presence of aligned circular cracks lead to a substantial anisotropy even at low crack densities. The considerable deviation from anisotropy in the presence of cracks is due to acute reduction in material stiffness along the crack normal compared to the transverse directions leading to anisotropy. Large deviation from isotropy even at low crack densities indicate that the presence of cracks have a much more significant effect on deviation from isotropy than pores. 

Analogous to overall anisotropy in particulate reinforced composite materials, homogeneous isotropic materials containing defects such as pores and cracks exhibit overall isotropy under certain conditions. Unlike particulate reinforced composite materials, only microstructural descriptors have an influence on overall isotropy in porous and cracked materials. Overall isotropy in homogeneous isotropic materials can be inferred by anisotropy indices $A^{U*}$, $A^{L*}$ and $A^{SE*}$ being consistently zero with a variation in microstructural descriptors such as pore volume fraction, shape and shape parameters. 

\begin{figure*}[htb]
    \begin{subfigure}[t]{0.4\textwidth}
    \includegraphics[width=\textwidth,trim= {0.0cm 4.5cm 1.0cm 5.6cm},clip]{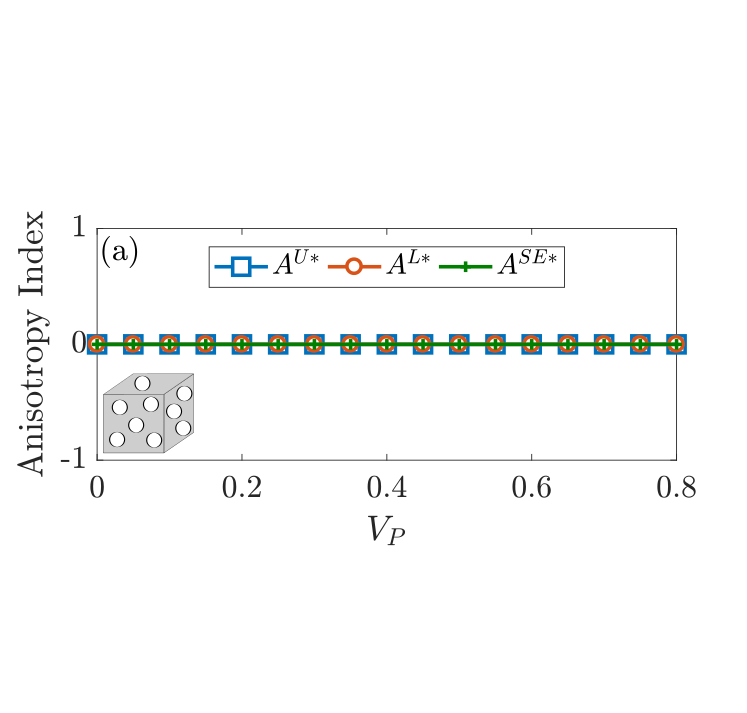}
    \phantomsubcaption
    \label{fig:spheres_pores}
    \end{subfigure} \qquad
    \begin{subfigure}[t]{0.4\textwidth}      
    \includegraphics[width=\textwidth,trim= {0.0cm 4.5cm 1.0cm 5.6cm},clip]{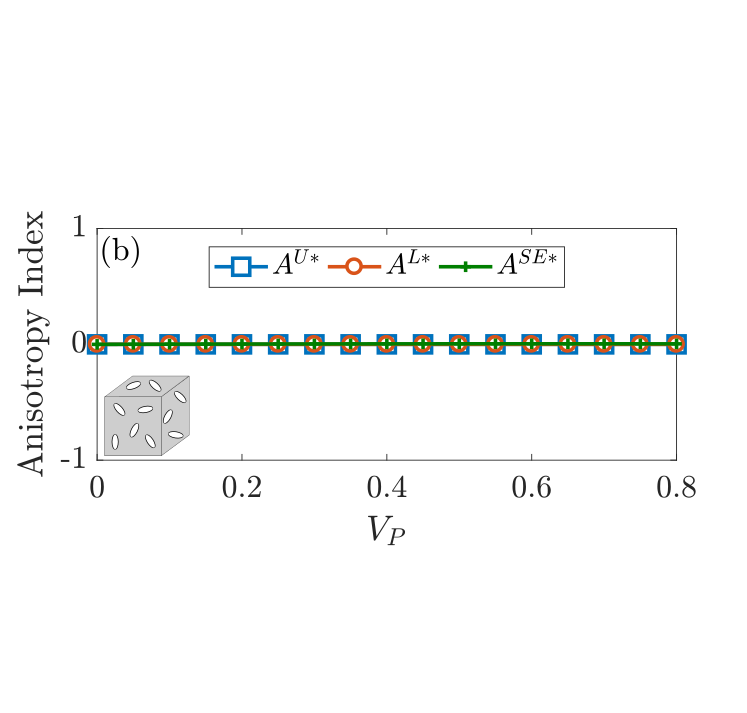}
    \phantomsubcaption
    \label{fig:RandomProlate_pores}
    \end{subfigure} \\
    \begin{subfigure}[t]{0.4\textwidth}      
    \includegraphics[width=\textwidth,trim= {0.0cm 4.5cm 1.0cm 5.6cm},clip]{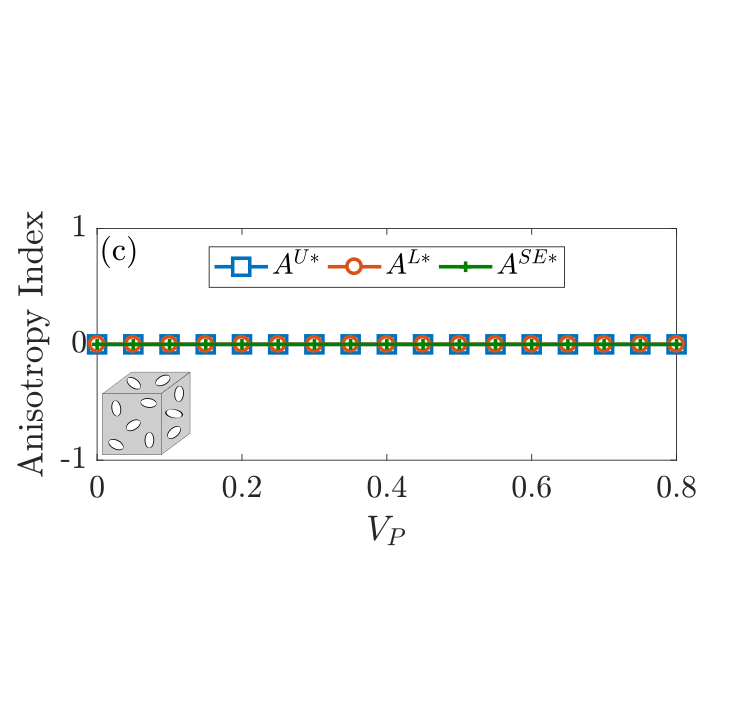}
    \phantomsubcaption
    \label{fig:RandomOblate_pores}
    \end{subfigure} \qquad
    \begin{subfigure}[t]{0.4\textwidth}      
    \includegraphics[width=\textwidth,trim= {0.0cm 4.5cm 1.0cm 5.6cm},clip]{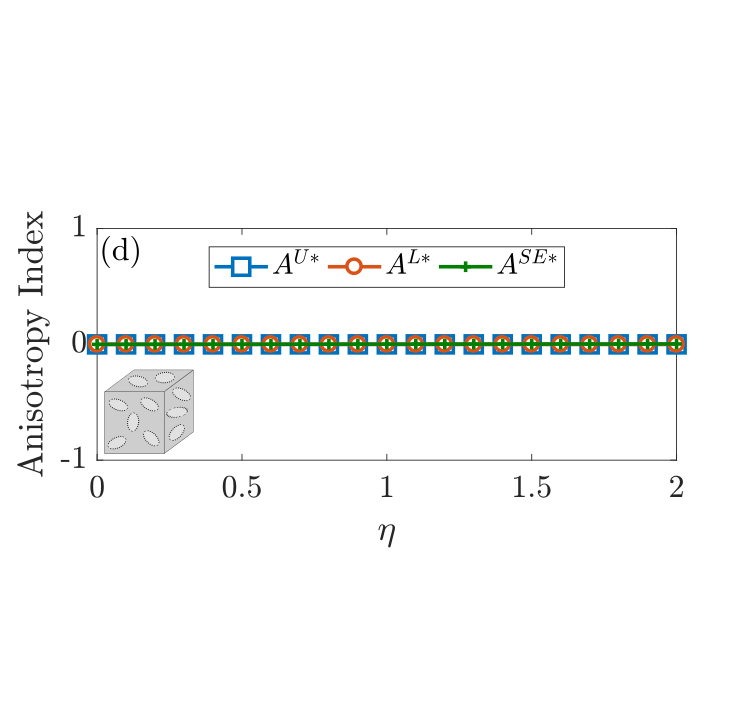}
    \phantomsubcaption
    \label{fig:RandomCracks}
    \end{subfigure}   
    \vspace{-10pt} 
    \caption{Isotropic homogeneous medium consisting of pores/cracks exhibit overall isotropy indicated by anisotropy indices $A^U$,$A^L$,$A^{SE}$ being zero under the following conditions (composite configuration mentioned in brackets): (a) pores are spherical, (b) randomly oriented prolate ellipsoids, ($\alpha=2$), (c) randomly oriented oblate ellipsoids, ($\alpha=1/2$), and (d) randomly oriented cracks.}
    \label{fig:anisotropy_6}
\end{figure*}

Porous homogeneous isotropic materials exhibit overall isotropy  when the pores are spherical (Fig.\ref{fig:spheres_pores}) since spheres have identical response in all directions. Additionally, randomly oriented ellipsoidal pores lead to overall isotropy. Random distribution of pores can be modeled by co-ordinate transformation of the strain concentration tensor and averaging the transformed strain concentration tensor over all possible orientations analogous to the case of randomly oriented particulate composites (Eq. \ref{eq:AMT}-\ref{eq:MTrot}). The effective elastic stiffness with randomly oriented pore phase ($C^{*}_{Pore-Random}$) can be evaluated considering the orientation averaged strain concentration tensor in the following form.

\begin{equation}
 C^{*}_{Pore-Random}=(1-V_P ) C^m  A^{MT}_{Random}
\label{eq:moritanakaaveragedpore}
\end{equation}

Evaluation of $A^{U*}$, $A^{L*}$ and $A^{SE*}$ considering $C^{*}_{Pore-Random}$ corresponding to randomly oriented ellipsoidal pores reveal that they are zero across the range of practical pore volume fractions  as shown in Fig. \ref{fig:RandomProlate_pores} and Fig. \ref{fig:RandomOblate_pores}) indicating overall isotropy irrespective of overall shape. Similarly, randomly oriented circular cracks also exhibit overall isotropy over a range of crack density $\eta$ indicated by the anisotropy indices being zero as shown in Fig. \ref{fig:RandomCracks}. Previous observations regarding overall isotropy in porous and cracked materials can be rigorously proved with the current approach to quantify anisotropy in heterogeneous materials.

\subsection{Elastic anisotropy in fiber reinforced composite materials}

Fiber reinforced composite materials represent a class of composite materials that are relevant for engineering applications due to their high specific stiffness and strength. Isotropic matrix reinforced with isotropic unidirectional continuous fiber reinforcement leads to transversely isotropic composite materials requiring 5 independent constants to fully populate the stiffness matrix. Although the choice of the independent constants are the The effective elastic constants of a unidirectional fiber reinforced composite material consisting of isotropic fibers ($E_f$,$\nu_f$) embedded in isotropic matrix ($E_m$,$\nu_m$) can be obtained by a combination of composite cylinder assembly and Generalized self consistant scheme in terms of axial Young's modulus $E_{11}^C$, axial Poisson's ratio $\nu_{12}^C$, axial shear modulus $G_{12}^C$, transverse shear modulus $G_{23}^C$, and transverse bulk modulus $K_{23}^C$.

The stiffness matrix of a two phase unidirectional continuous fiber reinforced composite material can be expressed in the following form.
\begin{widetext}
\begin{equation}
\label{eq:cca}
  {
C^*_{UD-FRC}  = 
 \begin{bmatrix}
   E_{11}^C +4 (\nu_{12}^C)^2 K_{23}^C &  2 \nu_{12}^C K_{23}^C & 2 \nu_{12}^C K_{23}^C &  0 & 0 & 0 \\
   2 \nu_{12}^C K_{23}^C &  K_{23}^C+G_{23}^C & K_{23}^C-G_{23}^C &  0 & 0 & 0 \\
   2 \nu_{12}^C K_{23}^C &  K_{23}^C-G_{23}^C & K_{23}^C+G_{23}^C &  0 & 0 & 0 \\
   0 &  0 & 0 &  G_{23}^C & 0 & 0 \\
   0 &  0 & 0 &  0 & G_{12}^C & 0 \\
   0 &  0 & 0 &  0 & 0 & G_{12}^C \\
 \end{bmatrix} 
  } 
\end{equation}
\end{widetext}

where the independent constants $E_{11}^C$, $\nu_{12}^C$, $G_{12}^C$, $G_{23}^C$, and $K_{23}^C$ are evaluated from a combination of composite cylinder assembly approach and generalized self consistant scheme \cite{zhang2014micromechanics}. While composite cylinder assembly approach provides closed form expressions for $E_{11}^C$, $\nu_{12}^C$, and $G_{12}^C$, generalized self consistant scheme yields $G_{23}^C$ and $K_{23}^C$  from the roots of a quadratic governing equation \cite{zhang2014micromechanics}.

\begin{figure*}[htb]
    \centering
    \begin{subfigure}[t]{0.4\textwidth}
    \includegraphics[width=\textwidth]{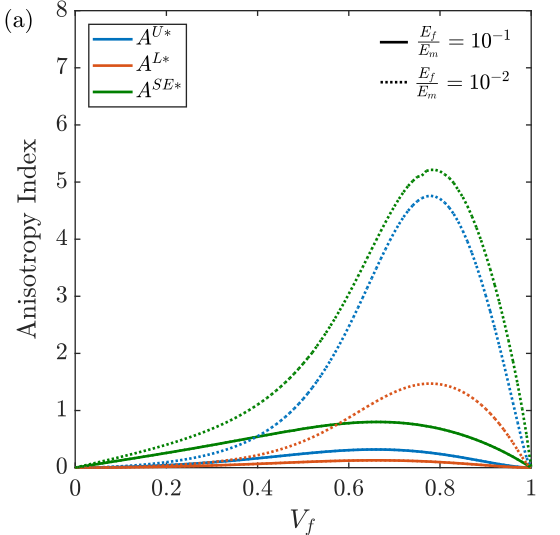}
    \phantomsubcaption
    \label{fig:anisotropy_CCA1}        
    \end{subfigure} \qquad
    \begin{subfigure}[t]{0.4\textwidth}
    \includegraphics[width=\textwidth]{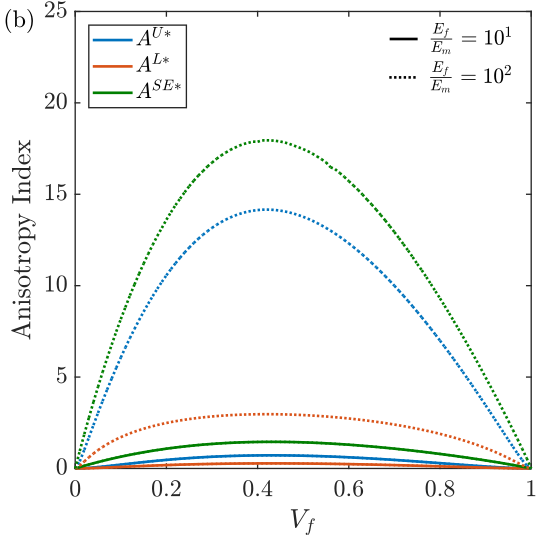}
    \phantomsubcaption
    \label{fig:anisotropy_CCA2}
    \end{subfigure}  
    \begin{subfigure}[t]{0.4\textwidth}      
    \includegraphics[width=\textwidth]{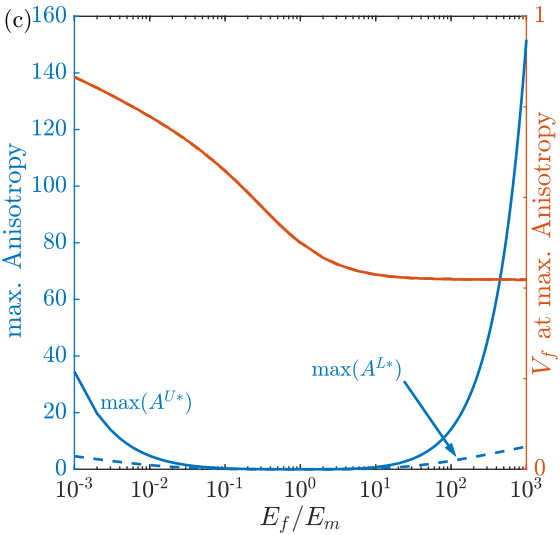}
    \phantomsubcaption
    \label{fig:anisotropy_CR_effect1}
    \end{subfigure} \qquad
    \begin{subfigure}[t]{0.4\textwidth}      
    \includegraphics[width=\textwidth]{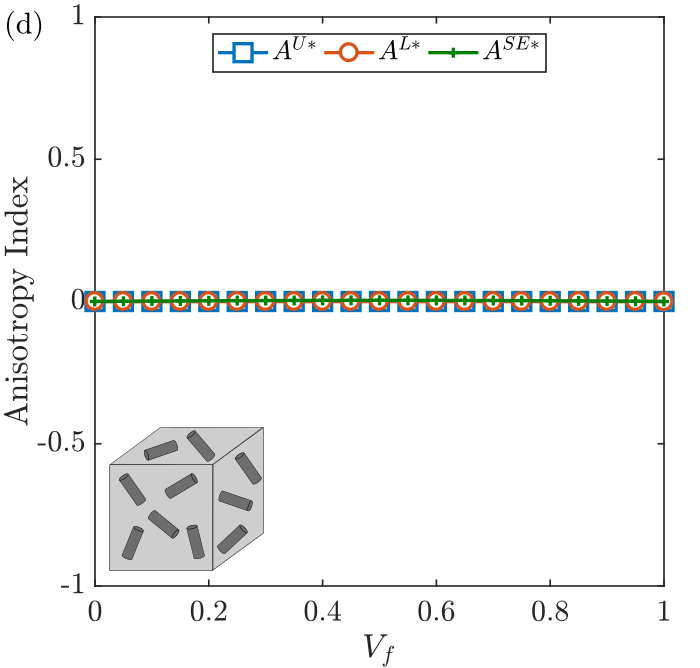}
    \phantomsubcaption
    \label{fig:anisotropy_CR_effect2}
    \end{subfigure}
    \vspace{-10pt}      
    \caption{Influence of microstructural and material descriptors on anisotropy in two phase unidirectional fiber reinforced composite materials: effect of (a) fiber volume fraction with compliant fibers ($E_f/E_m<1$), (b) fiber volume fraction with stiffer fibers ($E_f/E_m>1$), (c) maximum anisotropy and particulate volume fraction at which maximum anisotropy occurs for unidirectional fiber reinforced composite materials, and (d) overall isotropy in randomly oriented short fiber composites indicated by anisotropy indices $A^U$,$A^L$,$A^{SE}$ being zero ($E_f/E_m=2$).}
    \label{fig:anisotropy_7}
\end{figure*}

Anisotropy quantification in unidirectional continuous fiber reinforced composites utilizing the effective elastic stiffness $C^*_{UD-FRC}$ ( Eq. \ref{eq:cca} ) has been summarized in Fig. \ref{fig:anisotropy_7}. The nature of anisotropy variation with fiber volume fraction exhibits some similarities to the case of particle reinforced composite materials described previously. The anisotropy indices increase from zero ($V_f=0$), reach a peak and reduce back to zero ($V_f=1$) highlighting heterogeneity induced anisotropy. However, fibers more compliant than the matrix ($E_f/E_m<1$) and stiffer than matrix ($E_f/E_m<1$) exhibit a differences in the nature of anisotropy variation with fiber volume fraction as shown in Fig. \ref{fig:anisotropy_CCA1}, \ref{fig:anisotropy_CCA2} highlighting the  role of elastic contrast in influencing the nature of anisotropy variation apart from fiber volume fraction in fiber reinforced composite materials. 

Variation of peak anisotropy and fiber volume fraction at which peak anisotropy occurs has been presented in Fig. \ref{fig:anisotropy_CR_effect1}. While peak anisotropy increases with large elastic mismatch between the fiber and matrix, i.e $E_f << E_m$ and $E_f>>E_m$ it remains close to zero implying minor deviation in isotropy in the range $10^{-1}<E_f/E_m<10^1$. The fiber volume fraction corresponding to peak anisotropy reduces with increasing elastic contrast becoming asymptotic in the vicinity of $V_f=40\%$ beyond $E_f/E_m=10$. This behavior, in contrast to the 
volume fraction corresponding to peak anisotropy exhibiting a minima 
in the vicinity of $E_f/E_m\approx1$ in particle reinforced composites, illustrates the influence of secondary phase shape in influencing peak anisotropy.

Analogous to the microstructural descriptors leading to overall isotropy in two phase particulate reinforced composite materials, fiber reinforced composite materials consisting of randomly oriented isotropic short fibers embedded in isotropic matrix exhibit overall isotropy as shown in Fig. \ref{fig:anisotropy_CR_effect2}.  

Anisotropy quantification considering aligned and random orientation of particulate reinforcement, homogeneous materials containing pores/cracks and fiber reinforced composites demonstrate the generalizability of the present approach to consider any family of heterogeneous material with effective elastic stiffness evaluated by appropriate homogenization technique. This enables anisotropy quantification in realistic heterogeneous material microstructures consisting of multiple secondary phases, multiple families of pores, and cracks where each constituent phase can be individually anisotropic. 

While $A^{U*}$, $A^{L*}$ and $A^{SE*}$ can be utilized to quantify anisotropy in heterogeneous materials, the most suitable index for anisotropy quantification is a pertinent question. $A^{U*}$ and $A^{L*}$ have simple closed form expressions in terms of the effective stiffness and compliance matrices of a heterogeneous material leading to ease of evaluation but their physical meaning is difficult to comprehend. $A^{SE*}$ on the other hand requires optimization which is computationally expensive and sensitive to the optimization parameters. However, $A^{SE}$ has a clear physical meaning and has an advantage over $A^{U*}$ and $A^{L*}$ in the form of easy generalization to consider physical fields other than mechanical loading, multi-physics phenomenon and non-linearity. These factors drive the selection of the appropriate anisotropy index. 

The current approach to quantify anisotropy in heterogeneous materials along with the presented results open up the exploration of the role of salient microstructural features and microstructure statistics in influencing anisotropy. In random heterogeneous materials, microstructural features such as phase topology and connectivity also play a role in influencing anisotropy that can be quantified and correlated with the microstructure with the current approach. Further, the current approach for anisotropy quantification can form the basis for anisotropy tuning in composite \cite{vidyasagar2018microstructural} and architectured materials \cite{kumar2020inverse,zheng2021data} leading to the design of materials with desired anisotropy.

\section{Conclusion}

In conclusion, this work presents a novel approach to quantify anisotropy in heterogeneous materials accounting for material and microstructural descriptors of heterogeneity. The proposed approach  
reinterprets single crystal anisotropy indices considering the effective elastic properties evaluated by appropriate homogenization technique. Anisotropy quantification in two-phase composite materials reveals that heterogeneity with elastic mismatch between the constituent phases induces anisotropy. The influence of material and microstructural descriptors on anisotropy in heterogeneous materials is a significant distinction from single crystals. Variation of maximum anisotropy with material and microstructural descriptors indicate a clear threshold volume fraction above which peak anisotropy occurs. Specific material and microstructural descriptors leading to overall isotropy in two phase composite materials and extension of the approach to consider porosity and cracks have also been presented demonstrating the applicability of the approach to any class of heterogeneous materials. The presented approach and results opens up the exploration of salient microstructral features and microstructural statistics in influencing anisotropy in heterogeneous, porous and architectured materials. Additionally, the proposed approach can be extended to consider other physical fields, multiphysics phenomena, and nonlinearity in heterogeneous materials.


\vskip2pc

\bibliographystyle{unsrt}
\bibliography{RSPA_Author_tex}
\end{document}